\documentclass[sigconf,screen,nonacm]{acmart}

\AtBeginDocument{%
  }

\usepackage{graphicx}
\usepackage{colortbl}
\usepackage{subcaption}
\usepackage{makecell}
\usepackage{csquotes}
\usepackage{enumitem} 

\usepackage{arydshln}
\newcommand{\rowcollight}{\rowcolor{gray!10}}

\PassOptionsToPackage{table}{xcolor}
\usepackage{xcolor}
\usepackage{graphicx}  
\usepackage{colortbl}
\usepackage{subcaption}
\usepackage{makecell}
\usepackage{enumitem}
\usepackage{multirow}

\definecolor{mygray}{gray}{0.9}

\begin{document}

\title{Designing Beyond Language: Sociotechnical Barriers in AI Health Technologies for Limited English Proficiency}

\author{Michelle Huang}
\orcid{0000-0002-5958-5942}
\affiliation{%
  \institution{University of Illinois Urbana-Champaign}
  \city{Urbana}
  \state{Illinois}
  \country{USA}
}
\email{mh106@illinois.edu}

\author{Violeta J. Rodriguez}
\orcid{0000-0001-8543-2061}
\affiliation{%
  \institution{University of Illinois Urbana-Champaign}
  \city{Urbana}
  \state{Illinois}
  \country{USA}}
\email{vjrodrig@illinois.edu}

\author{Koustuv Saha}
\orcid{0000-0002-8872-2934}
\authornote{Both authors are advisors of this work.}
\affiliation{%
  \institution{University of Illinois Urbana-Champaign}
  \city{Urbana}
  \state{Illinois}
  \country{USA}}
\email{ksaha2@illinois.edu}

\author{Tal August}
\authornotemark[1]
\orcid{0000-0001-6726-4009}
\affiliation{%
  \institution{University of Illinois Urbana-Champaign}
  \city{Urbana}
  \state{Illinois}
  \country{USA}}
\email{taugust@illinois.edu}

\renewcommand{\shortauthors}{Huang et al.}

\begin{abstract}
    Limited English proficiency (LEP) patients in the U.S. face systemic barriers to healthcare beyond language and interpreter access, encompassing procedural and institutional constraints. AI advances may support communication and care through on-demand translation and visit preparation, but also risk exacerbating existing inequalities. We conducted storyboard-driven interviews with 14 patient navigators to explore how AI could shape care experiences for Spanish-speaking LEP individuals. We identified tensions around linguistic and cultural misunderstandings, privacy concerns, and opportunities and risks for AI to augment care workflows. Participants highlighted structural factors that can undermine trust in AI systems, including sensitive information disclosure, unstable technology access, and low literacy. While AI tools can potentially alleviate social barriers and institutional constraints, there are risks of misinformation and reducing human-to-human interactions. Our findings contribute AI design considerations that support LEP patients and care teams via rapport-building, educational and language support, and minimizing disruptions to existing practices.\footnote{Supplementary materials available at \url{https://github.com/mhuang412/designing-beyond-language}}
\end{abstract}



\maketitle

\section{Introduction}

Patient-provider communication is strongly linked to health outcomes and quality of care~\cite{arora2003interacting, street2009does, ong2000doctor}. Effective communication builds therapeutic relationships, satisfaction, trust, and long-term health outcomes~\cite{street2009does, ha2010doctor}, while poor communication contributes to errors, compromised patient safety, and inefficient use of resources~\cite{vermeir2015communication, tiwary2019poor}. Language barriers in particular can exacerbate these challenges. In the U.S., over 25 million individuals aged 5 years and older report speaking English less than ``very well''~\cite{Census}. These individuals with limited English proficiency (LEP) are more likely to experience misdiagnoses, medication complications, poorer adherence to treatment plans, and decreased comprehension of diagnoses~\cite{crane1997patient, gandhi2000drug, flores2006language, tiwary2019poor}. Consequently, patient-provider communication, relationship-building, and shared decision-making break down~\cite{paredes2018influence, flores2006language}, reducing both patient and provider satisfaction~\cite{al2020implications}. 

Digital health technologies, such as telehealth, online patient portals, and mobile health applications, can facilitate asynchronous communication and on-demand information access, which in principle could improve the quality and convenience of healthcare by enabling continuous and accessible support~\cite{fatehi, madanian, ezeamii}. However, LEP populations underutilize these resources due to persistent implementation barriers such as low digital and health literacy, unreliable Internet access, and ethical and privacy concerns~\cite{ahmed, khan, madanian, nouri2020addressing, Payvandi}. Even technologies designed specifically to accommodate LEP individuals, such as multilingual patient portals, are often poorly integrated or have limited functionalities, thus discouraging use~\cite{Anthony}. 

Emerging AI-powered health technologies have the potential to provide personalized and timely medical support and translation~\cite{johnson, genovese, miller}, enabling LEP patients to engage with the healthcare system more effectively. For example, on-demand AI translation can facilitate patient-provider conversations~\cite{mehandru-etal-2023-physician}, and new large language model (LLM)-powered chatbots can summarize and simplify complex medical information~\cite{guo2021automated} or help patients prepare for visits~\citep{10.1145/3706598.3714196, 10.1145/3715336.3735674}. However, most AI-powered interventions target digitally literate, English-speaking patients \citep{jin2024better, naamati2024beyond}, and the underlying models generally struggle to adapt to marginalized cultural \citep{santy-etal-2023-nlpositionality} and language backgrounds \cite{10.1145/3715275.3732045, sap-etal-2019-risk}. Furthermore, little is known about how AI interventions should be designed and governed when language barriers intersect with cultural norms, privacy fears, and low literacy, especially in contexts where human intermediaries (e.g., patient navigators) already play a critical role. Therefore, the rapid deployment of such AI health interventions risks replicating or further exacerbating existing disparities~\cite{ahmed}. 

In this paper, we seek to understand the risks and opportunities that AI health interventions pose for LEP patients, guided by the following research questions (RQs):

\begin{enumerate}
    \item [\textbf{RQ1:}] How can linguistic, cultural, or sociotechnical factors shape how LEP patients perceive health technologies?
    \item[\textbf{RQ2:}]   
    How can AI health technologies facilitate culturally responsive, trustworthy, and effective communication and collaboration between LEP patients and healthcare teams?
\end{enumerate}

To address the above RQs, we conducted 14 storyboard-centered interviews with individuals who support Spanish-speaking LEP individuals receiving care in the U.S. These individuals, whom we collectively refer to as \textit{patient navigators}, assisted patients in various ways, including accompanying them to provider visits, translating, teaching digital skills, and bridging resources. Their role as mediators between patients and providers lends them insights into the broader contexts and relational dynamics that shape LEP patient experiences. Consequently, they build on their experiences to reason about the feasibility and risks of developing AI technologies to facilitate equitable communication and care.

From these interviews, we identified key communication and systemic challenges that Spanish-speaking LEP patients face, such as low literacy and privacy concerns. These challenges also highlighted possible risks and opportunities for using AI, such as reducing social barriers to care while potentially diminishing existing human support. Our findings underscore the importance of adapting to cultural and linguistic nuances, overcoming various types of literacy barriers, and respecting user privacy to foster trust. These findings inform practical guidelines for developing AI tools that support LEP populations. Our contributions are summarized as follows:

\begin{itemize}
\item First, we illuminate cultural, communication, and structural challenges faced by Spanish-speaking LEP patients, their providers, and patient navigators. 
\item Second, we identify both perceived risks and opportunities for AI to support patient experiences. 
\item Third, we propose design guidelines and opportunities for AI tools that support patient access to healthcare resources and mediate culturally sensitive interactions.
\end{itemize}

Together, these contributions aim to inform when to---and when not to---develop AI health interventions and how to do so in a way that is attuned to the needs of linguistically and culturally diverse patient populations.

\section{Related Work}

\subsection{Mediating Clinical Communication with LEP Individuals}
Effective healthcare requires two-way communication in which providers contribute medical expertise and patients share their needs and preferences~\cite{iroegbu2025influence, williams2017patient}. Strong interpersonal communication has been shown to influence patient outcomes as much as, if not more than, clinical instruction~\cite{bartlett1984effects}, leading to improved adherence to treatment plans~\cite{street2009does}, understanding of diagnoses~\cite{kennedy2014improving}, and satisfaction~\cite{ruben2020communication}. Conversely, impeded communication is linked to more errors~\cite{tiwary2019poor}, worsened decision-making~\cite{paredes2018influence}, and lower rates of physician visits and preventive services~\cite{jacobs2006need, woloshin1997language, fiscella2002disparities}. For LEP individuals, these challenges are compounded by linguistic~\cite{schouten2020mitigating}, cultural~\cite{lazaro2024literacy}, intersectional~\cite{parthab2013cross}, and procedural~\cite{frank2000medical} barriers. Therefore, there is growing interest in improving health outcomes by studying communication strategies that integrate patients' lived experiences and values~\cite{seo2021challenges, 10.1145/3706598.3713782}.

Research shows that language-concordant care, whether via direct communication ~\cite{molina2019power} or interpretation ~\cite{dunlap2015effects, eskes2013patient}, leads to fewer communication errors~\cite{karliner2007professional} and higher satisfaction and perceived quality of care~\cite{diamond2019systematic} for LEP individuals. Despite these benefits, interpretation services remain underutilized due to barriers such as limited availability, cost, time constraints, and policies~\cite{khoong2021addressing, jaeger2019barriers, jacobs2004overcoming}. Even when interpretation is available, institutional policies can constrain interpreters to verbatim translation rather than clarification or advocacy, leaving unresolved misunderstandings and limiting relationship-building—conditions that shape what AI can realistically improve. Beyond linguistic access, broader sociocultural factors complicate LEP communication. Patients may self-censor due to stigma, fear of judgment, or mistrust of institutions~\cite{levy2018prevalence, el2023medical}. These dynamics highlight the need for cultural sensitivity, empathy, and attention to the broader contexts of patients' lives. Patient navigators and community health workers (CHWs) function as critical infrastructure for marginalized communities, supporting interpretation, care coordination, documentation, and advocacy~\cite{khatri2024enablers}. Prior work highlights how these roles reduce access barriers but also create invisible labor and emotional burdens~\cite{verdezoto2021invisible}. Thus, understanding how AI might support or destabilize this ecosystem is necessary for equitable deployment.

\subsection{AI Tools in Healthcare}
Recent AI advancements have potential to support healthcare systems by improving clinical decision-making and workflows, resource allocations, medical imaging, patient monitoring, medical translations, and patient-provider relationships~\cite{maleki2024role, shaheen2021applications, barwise2024using, koski2021ai}. However, prior work has also identified key challenges and risks when implementing AI for healthcare, including data privacy, lack of transparency in decision-making, misinformation, bias mitigation, and cultural sensitivity~\cite{maleki2024role, germani2024dual, hussain2024humanizing, capraro2024impact, chustecki2024benefits,shi2025mapping,yoo2025ai}. 

AI has potential to support marginalized communities in receiving care by overcoming language barriers, providing personalized health information, and offering support in rural or low-resource areas~\cite{genovese, kaur2025enhancing}. \citet{barwise2024using} explored patients' perception of using predictive models for interpreter prioritization and \citet{mehandru-etal-2023-physician} used physician-in-the-loop methods to improve reliance of machine translation in settings where no human translators were available. Other work has also explored using chatbots for pre-clinical consultations \citep{10.1145/3706598.3714196, 10.1145/3715336.3735674} and helping patients read physician notes \citep{kambhamettu2025traceable}, discharge instructions \citep{zaretsky2024generative}, and medical information \citep{guo2021automated}. While AI presents many exciting possibilities to support patient care, the majority of such interventions are focused on English or translating into English, posing issues for low-resource languages such as reinforcing biases and overlooking cultural and infrastructural challenges that may cause AI to exacerbate rather than alleviate inequities in underserved populations~\cite{wasi2025ai, yuan2023adoption, iloanusi2024ai}. The LLMs driving much of current AI developments are known to encode biases on race \citep{sap-etal-2019-risk, hofmann2024ai}, gender \citep{ding2025gender}, language \citep{alhanai2025bridging}, and culture \citep{santy-etal-2023-nlpositionality}. While these challenges and biases are known, there is work needed to determine if and how AI can be designed for LEP communities while taking into consideration current model limitations and existing healthcare support systems.

\subsection{Considerations for Health Technologies in Marginalized Communities}
While digital health technologies generally expand access to care~\cite{raney2017digitally, al2022improving,suh2024rethinking}, they are often underutilized by LEP patients~\cite{nouri2020addressing, gordon2025lower, hsueh2021disparities}, thus exacerbating existing inequalities~\cite{yao, linggonegoro2021telemedicine}. For example, during and after the COVID-19 pandemic, LEP patients utilized video visits and online patient portals at a lower rate than English-proficient patients~\cite{rodriguez2021differences}. 

Several structural factors impede health technology access and utilization for LEP individuals. Limited reading, digital, and health literacy can restrict meaningful engagement with digital tools~\cite{lazaro2024literacy, smith2019new}, yet digital health literacy is key to realizing equitable advancements in healthcare~\cite{bywall2025promoting}. Prior work in HCI also highlights the need for AI technologies to be culturally sensitive~\cite{10.1145/3706598.3713362, 10.1145/3449239} in order to be more trustworthy and accessible to marginalized users. Furthermore, LEP populations face generally lower rates of access to technology~\cite{higashi2025digital}, but even when access is available, they report distrust in digital services and difficulty understanding privacy policies~\cite{tan2022addressing}, fueling their hesitation in adopting technology that solicits their personal information. Recent legal actions and policies in the U.S. have compounded these concerns~\cite{page2017chilling, fleming2019us}, underscoring the need to prioritize privacy and trust in equitable technology design. There are, however, community-based health service approaches that have proven to be accessible and productive for LEP individuals. SMS programs, WhatsApp groups, and CHW-linked tools used by Latino immigrant communities have been shown to support self-management and information access~\cite{aibo2024community, anaya2021meeting}.

In consideration of the barriers, benefits, and risks for technology to support LEP individuals in healthcare, it is important to understand how these technologies can be implemented equitably and safely~\cite{10.1145/3706598.3713165}. There is little recent work that explores all of these factors together for LEP populations in particular. Accordingly, we report our findings around the barriers and experiences that LEP individuals may face when receiving healthcare, and in what ways AI technology can be designed to facilitate communication with their healthcare teams.

\section{Methods}

\subsection{Study Context}

In this study, we interviewed patient navigators who support Spanish-speaking LEP individuals when interacting with healthcare systems in the U.S. We focused on patient navigators, rather than patients themselves, because as we reflected on the tensions involved in conducting research with a marginalized community (in this case, LEP individuals seeking care), we were concerned that starting with LEP individuals might risk perpetrating further harm to the community by taking data and information and providing little in return \citep{liang2021embracing}. Speaking directly with patients at the outset could also risk misrepresenting or excluding key aspects of their experiences if we failed to frame the right questions or if patients were hesitant to disclose sensitive information, especially since the lead researcher in the study was not themselves a part of this community (\autoref{positionality}). In contrast, patient navigators routinely accompany many patients through the care-seeking process, providing crucial perspectives that surface both recurring experiences and specific incidents that reveal how patients engage with healthcare and technology. Talking to navigators allowed us to gain an understanding of the culture and background of LEP individuals, identify potential avenues for giving back to the community (e.g., presenting at community events, conducting follow-up design workshops), and build relationships with these communities for further research into technologies that are sensitive to the needs and lived experiences of LEP individuals. We recognize that navigators themselves may also be part of LEP communities or other marginalized groups. Accordingly, we worked closely with three navigators to provide feedback on pilot studies and assess what value such a study might bring to navigators. We discuss future work building on our findings to bring benefits of new AI health tools to both LEP individuals and patient navigators (\autoref{sec:discussion}).

Initial feedback from navigators was that LEP individuals vary widely in their care experiences due to the range of languages, backgrounds, and cultures. In response to this feedback, we focused on Spanish-speaking LEP individuals for two reasons: Spanish is by far the most commonly spoken language at home after English in the U.S.~\cite{Census}, and the majority of organizations we worked with focused primarily on Spanish speakers, giving us greater access to navigators and community resources with this population. However, as we will further discuss in our findings (\autoref{subsec:findingsLanguage}), even with a shared language, individuals had varying experiences and identities (e.g., be immigrants from various countries or U.S.-born). In the rest of this paper, we refer to Spanish-speaking LEP individuals as just \textit{LEP individuals}, for simplicity.

\subsection{Study Design}

\subsubsection{Interest survey}
\label{subsec:interestSurvey}
Interested patient navigators completed an online screening survey that asked for demographic information, including age range, gender, ethnicity, occupation, organization, years of relevant experience in their roles, and languages spoken. We also adapted a 5-question basic AI literacy questionnaire \cite{grassini2024psychometric} in the interest form (see \autoref{ailiteracy}). After completing the survey questions, participants signed up for a 60-minute interview within the next week. A copy of the consent form was included for the participants to review and sign before the interview.

\subsubsection{Storyboard interviews}
We conducted 60-minute semi-structured storyboard interviews over Zoom with patient navigators who filled out the survey. All interviews were conducted in English. Participants were compensated with a \$20 Amazon gift card at the completion of the study. This study was approved by our organization's Institutional Review Board. 

Each interview began with questions about the navigator's background, job responsibilities, and the characteristics of the patients they typically supported. We also asked about challenges they face in their role and challenges their patients face in order to identify pain points in current practices and relationships between Spanish-speaking LEP patients, navigators, and providers as opportunities for AI intervention. We then extended these discussions to consider possible outcomes of integrating AI tools in patient care. 

Participants exhibited a range of knowledge on AI and ways in which they interacted with LEP individuals. To establish common ground across participants~\cite{davidoff2007rapidly}, we structured our interviews around six pre-defined storyboards. Storyboards can solicit immediate reactions and underlying needs probed by the hypothetical scenarios depicted~\cite{10.1007/978-3-030-23204-7_14, davidoff2007rapidly,kawakami2023sensing}. These storyboards surfaced participants' reactions and concerns towards potential AI systems, use cases, and patient responses that they might not otherwise think of. We explored the use of AI systems in various stages of care (e.g. before, during, and after provider visits) and in various capacities (e.g. translating, explaining, comforting).

For each interview, we screen-shared a slide deck with images of the storyboards. Participants were informed that the goal of the study was to understand their experiences with patients as a patient navigator and their perceptions of potential AI technologies. Storyboards were randomly ordered for each participant to control for potential bias introduced by ordering effects. For each storyboard, we asked if the storyboard was relatable to them and what their immediate reactions were. Then, follow-up questions were asked based on their responses or drawn from a semi-structured interview guide (available in the supplementary materials) as needed. Follow-up questions included: \textit{What are challenges that patients face during provider visits? How do patients generally seek medical information? How often do patients consider data privacy, whether online or offline?}

\begin{table*}[h]
\sffamily
\small
  \caption{Storyboard themes with motivating questions and supporting literature.}
  \label{tab:storyboardthemes}
  \Description{A table showing 6 rows for the storyboard themes around AI health technology development, and 3 columns for dimension, motivating ideas, and relevant literature}
  \begin{tabular}{lll}
    \textbf{Dimension} & \textbf{Motivating Ideas} & \textbf{Relevant Literature} \\
    \midrule
    Language \& Communication & \makecell[lt]{Can AI support accurate and productive conversations? \\ Can AI capture cultural and semantic nuances?} & \makecell[lt]{\citep{alhanai2025bridging}, \cite{antoniak2024nlp}, \cite{genovese}, \citep{guo2021automated}, \\ \citep{kambhamettu2025traceable}, \cite{10.1145/3706598.3714196}, \cite{mehandru-etal-2023-physician}, \citep{zaretsky2024generative}} \vspace{0.5mm}\\
    \arrayrulecolor{mygray}\hline
    Relationality \& Comfort & \makecell[lt]{Can patients interact with AI comfortably? \\ Can AI cater to the patient's specific needs and background?} &  \makecell[lt]{\cite{10.1145/3715275.3732045}, \cite{hofmann2024ai}, \cite{10.1145/3706598.3714196}, \cite{shaikh2024show},\\ \cite{wu2024aligning} } \vspace{0.5mm}\\
    \hline
    Clinical Safety \& Accuracy & \makecell[lt]{Can AI give accurate, safe, and appropriate medical information? \\ Are AI outputs biased or culturally insensitive?} & \makecell[lt]{\cite{10.1145/3628096.3628752}, \cite{hofmann2024ai}, \cite{jo2025understanding}, \cite{mehandru-etal-2023-physician},  \cite{10.1145/3715336.3735674}, \\ \citep{santy-etal-2023-nlpositionality},  \citep{sap-etal-2019-risk}, \cite{verran2024artificial}, \cite{zaretsky2024generative}} \vspace{.5mm}\\
    \hline
    Fairness, Equity, \& Access & \makecell[lt]{Does AI assume certain user privileges or characteristics? \\ Can AI worsen disparities?} & 
    \makecell[lt]{\cite{gordon2025lower}, \cite{higashi2025digital}, \cite{iloanusi2024ai}, \cite{kaur2025enhancing}, \\ \cite{lazaro2024literacy}, \cite{linggonegoro2021telemedicine}, \cite{wasi2025ai}, \cite{yuan2023adoption}} \vspace{.5mm}\\
    \hline
    Clinical Integration \& Support & \makecell[lt]{Can AI enhance clinical productivity? \\ How can AI support clinical staff safely and effectively?} & \makecell[lt]{\cite{10.1145/3628096.3628752}, \cite{antoniak2024nlp}, \cite{10.1145/3715336.3735674}, \cite{sezgin2023artificial}, \\ \cite{tierney2024ambient}, \cite{vijayakumar2023physicians}} \vspace{.5mm}\\
    \hline
    Privacy, Transparency, \& Ethics & \makecell[lt]{Do patients understand how their data will be handled? \\ Can AI alleviate or aggravate institutional issues or policies?} & \makecell[lt]{\cite{fleming2019us}, \cite{germani2024dual}, \cite{maleki2024role}, \cite{page2017chilling},\\ \cite{tan2022addressing}} \\
    \arrayrulecolor{black} \bottomrule
  \end{tabular}
\end{table*}

\subsection{Storyboard Construction}
Following prior health-focused studies in HCI~\cite{zhang2025knowledge, kawakami2023sensing}, we constructed the storyboards using an iterative method drawing from existing literature and our research team's expertise (\autoref{positionality}). We synthesized 34 papers across HCI, AI, and medical journals that propose use cases, risks, and benefits of AI in healthcare and for LEP individuals in particular. Then, a clinical psychologist and expert in health equity for under-served populations, in particular Spanish-speaking communities, reviewed proposed storyboards to ensure they were comprehensive, realistic, and would elicit meaningful discussions. We consolidated storyboards in six dimensions covering possible applications of AI health technologies for LEP patients (see \autoref{tab:storyboardthemes}). These dimensions included AI performance (linguistic nuances and reliability of information), ease of use (for patients and within clinical workflows), and systemic factors (access to technology and ethical concerns). Each dimension was used to create a 4-panel storyboard scenario that followed the same general structure: 1) patient encounters a challenge, 2) AI tool is introduced as a potential solution or support, 3) effects (either positive or negative) of AI emerge, and 4) patient reflects on the interaction with AI. Two example storyboards are shown in \autoref{fig:storyboard-ex}, and all storyboards are provided in the supplementary materials. A version of each storyboard with English translations was available for participants with limited Spanish proficiency.

All storyboards depict patients as the direct users of AI (except for the "Language \& Communication" theme in which a patient and navigator interact with AI together) because we initially centered this work around AI design considerations for LEP patients in isolation since this group is often left out of technology design. However, navigators later revealed barriers that complicate this reality, as further discussed in \autoref{subsec:disc1} and \autoref{subsec:disc4}. Furthermore, we intended for the scenarios to have a mix of positive and negative outcomes to encourage participants to consider the tensions between the benefits and risks of AI for LEP patient care. We strove to depict realistic AI-based current commercial or prototype systems (e.g. translation systems, chatbots). However, since storyboards inherently depict imagined rather than observed behavior, we intended to use them to elicit expectations, concerns, and situated judgments. This limitation is further discussed in \autoref{sec:limitations}.

\begin{figure*}[t]
    \centering
    \begin{subfigure}[t]{\textwidth}
        \centering
        \includegraphics[height=1.4in]{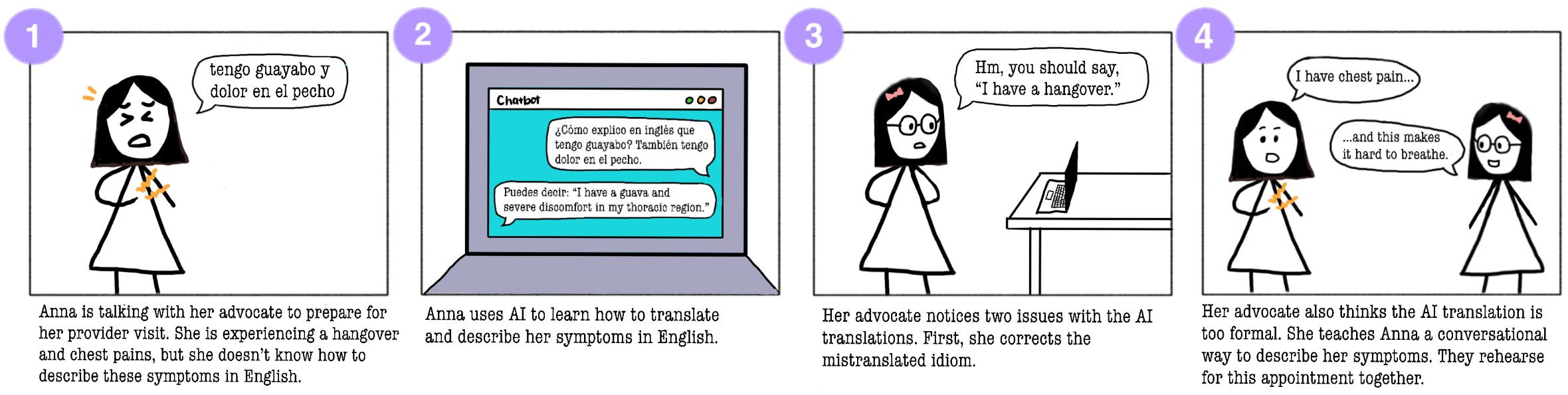}
        \caption{Language \& Communication theme: AI mistranslations of idioms and medical jargon showcase potential shortcomings.}
        \Description{A storyboard with four panels that show: 1) A woman is experiencing chest pains, 2) A chatbot interface on a computer screen shows a conversation in which the woman asks the chatbot in Spanish how to explain her symptoms in English, and the chatbot provides a translation, 3) A patient advocate notices and corrects mistranslations, 4) The woman and her patient advocate practice rehearsing her symptoms in English.}
    \end{subfigure}%

    \vspace{1em}
     
    \begin{subfigure}[t]{\textwidth}
        \centering
        \includegraphics[height=1.418in]{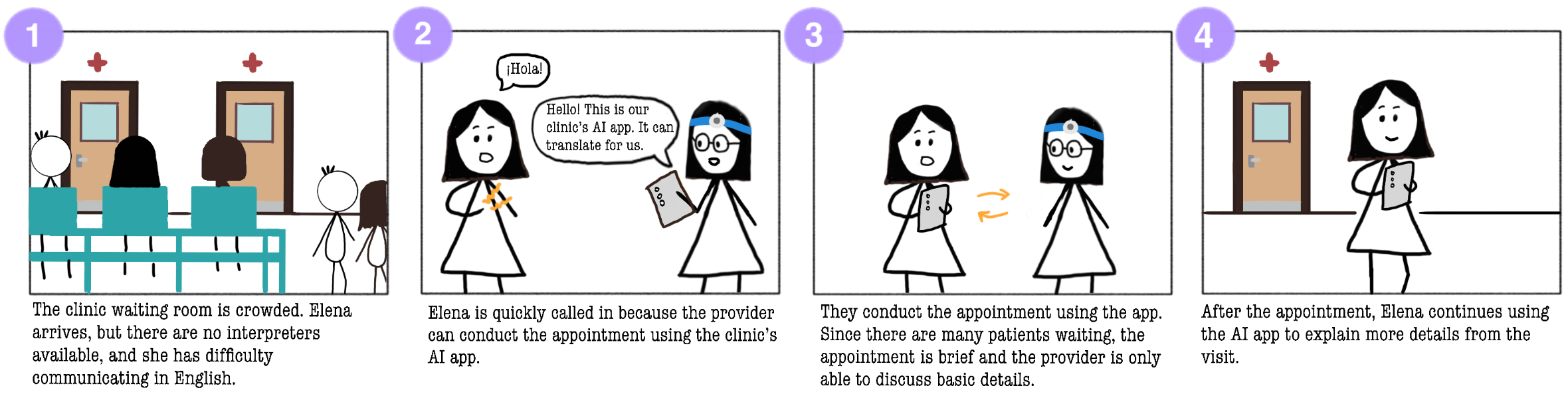}
        \caption{Clinical Integration \& Support theme: AI translator can support patient care when clinics are understaffed.}
        \Description{A storyboard with four panels that show: 1) A crowded clinic waiting room, 2) A Spanish-speaking patient is conversing with an English-speaking provider using an AI app to translate, 3) The patient and provider continue conversing through the AI translator, 4) The patient leaves the appointment and continues to use the AI app for more information.}
    \end{subfigure}
    
    \caption{Example storyboards used in our interviews.}
    \label{fig:storyboard-ex}
    \Description{Two 4-panel digitally-illustrated cartoon storyboards. The dialogue in the storyboards is in Spanish, but each panel has an associated English caption.}
\end{figure*}

\subsection{Recruitment}
\label{sec:recruitment}
We initially recruited navigators from two community non-profits that one of our research team members had prior connections with. We drafted the recruiting email and flyer that the point of contact at each organization distributed to their staff and organizational networks for voluntary sign-ups. These flyers and emails contained a link to the interest survey (\autoref{subsec:interestSurvey}). 

With the intention of gaining a holistic understanding of navigator and patient experiences, we aimed to interview patient navigators who worked across a diverse set of roles (e.g. interpretation, digital literacy, case management) and Spanish-speaking communities (e.g. children vs. adults, immigrants vs. U.S.-born, literate vs. illiterate). With this in mind, we also cold-emailed several other organizations in the region that provided health-adjacent services to Spanish-speaking populations. Eligibility criteria for scheduling an interview included being willing to participate in an interview, being able to communicate in English, and having experience directly interacting with Spanish-speaking patients in some capacity. We conducted interviews until data saturation was reached and no new information was collected (\autoref{sec:analysis}).

\subsection{Participants}
We recruited 14 patient navigators (8 female, 6 male) who supported Spanish-speaking LEP patients in the U.S. Participant roles included patient advocates, health liaisons, caseworkers, medical students, and interpreters. Participants came from eight distinct organizations, including medical schools, local community non-profits, and public school districts. Most of these organizations specifically serve, or at least have strong ties to, the Spanish-speaking or broader immigrant communities in their local areas. Therefore, patients either actively seek support from these organizations or are referred to them by word-of-mouth. The communities that these 14 participants work with generally consist of first- or second-generation immigrants from Hispanic countries. Because many of these LEP immigrants are undocumented, tightened immigration policies in the U.S. have led to greater institutional distrust and fear that disclosing their personal information may lead back to immigration enforcements. Therefore, navigators spent considerable efforts helping these individuals overcome barriers due to their status, such as applying for health insurance without a social security number.

These navigators also represent organizations with diverse medical care practices and resources. For example, some organizations allowed the use of ad hoc interpreters, while others required strictly professional interpreters. Some clinics had access to interpreters for low-resource languages like Q'anjob'al, while other clinics struggled to access even common languages like Spanish. Although we did not intentionally recruit patient navigators with similar cultural backgrounds as Spanish-speaking LEP individuals, 12 out of 14 participants reported speaking English and Spanish as their primary languages and having strong ties to the communities as first- or second-generation immigrants. We conducted interviews in English to align with participants’ professional working language and to reduce risks of misinterpretation in analysis. See \autoref{tab:participants} for a full breakdown of participants' demographics and roles.

Though not a requirement to participate in the study, participants also exhibited fairly high degrees of familiarity and proficiency with AI technologies. On a 5-point Likert scale ranging from 1 (strongly disagree) to 5 (strongly agree), participants reported understanding basic AI concepts $(M = 3.86, SD = 0.86)$ and being able to judge the benefits and risks of AI $(M = 3.64, SD = 0.74)$. All participants had extensive experience using AI tools such as ChatGPT or Google Translate in their work.

\begin{table*}
\sffamily
\footnotesize
    \setlength{\tabcolsep}{4pt}
  \caption{Summary of participants, including Occupation, Age, Gender, Race, Languages Spoken, and Years Working with Spanish-Speaking Patients (Yrs. w/ SSP).}
  \label{tab:participants}
  \Description{A table showing demographic information of 14 participants, with columns for participant, occupation, age range, gender, ethnicity/race, languages spoken, and years working with Spanish-speaking patients.}
  \begin{tabular}{cllclll}
    ID & Occupation & Age & Gender & Ethnicity/Race & Languages Spoken & \makecell{Yrs. w/ SSP} \\
    \toprule
    P1 & \makecell[tl]{Health Program Coordinator} & 25-34 & Woman & Hispanic/Latinx & English, Spanish & 1-3 years\\
   \rowcollight P2 & Patient Advocate & 18-24 & Woman & Hispanic/Latinx & English, Spanish & 3-5 years\\
    P3 & Caseworker & 35-44 & Man & Black/African American & English, Spanish, French & <1 year\\
    \rowcollight P4 & Community Worker & 18-24 & Woman & Black/African American & English, French, Lingala, Swahili & <1 year \\
    P5 & \makecell[tl]{Community Navigator Volunteer} & 25-34 & Woman & Hispanic/Latinx & English, Spanish & 5-10 years\\
    \rowcollight P6 & Home Visitor & 45-54 & Woman & White & English, Spanish & 15+ years\\ 
    P7 & Parent Liaison & 25-34 & Man & Hispanic/Latinx & English, Spanish & <1 years\\
   \rowcollight  P8 & Community Health Worker & 35-44 & Man & Hispanic/Latinx & English, Spanish & 15+ years\\
    P9 & Medical Student & 35-44 & Man & Hispanic/Latinx & English, Spanish & 3-5 years \\
   \rowcollight  P10 & Latino Family Liaison & 55-64 & Woman & Hispanic/Latinx & English, Spanish & 15+ years\\
    P11 & Medical Student & 18-24 & Man & Hispanic/Latinx \& White & English, Spanish & 3-5 years \\
    \rowcollight P12 & Digital Literacy Program Coordinator & 18-24 & Woman & Hispanic/Latinx & English, Spanish & 1-3 years\\
    P13 & Social Worker & 45-54 & Woman & Hispanic/Latinx & English, Spanish & 15+ years\\
   \rowcollight  P14 & Digital Literacy Program Coordinator & 18-24 & Man & Hispanic/Latinx & English, Spanish & 1-3 years\\ 
    \bottomrule
  \end{tabular}
\end{table*}

\subsection{Data Collection \& Analysis}
\label{sec:analysis}
All interviews were conducted, recorded, and transcribed via Zoom. One author reviewed each transcript for accuracy, removed pauses and filler words, and paraphrased quotations cited in this paper. We followed Braun and Clark's reflexive thematic analysis approach \cite{braun2006using, braun2019reflecting}. To familiarize ourselves with the data, two authors coded a subset of the interview transcripts independently. Each researcher identified quotes that were associated with our research questions and grouped these quotes with short descriptions. The two authors met to share initial ideas of interest and quotes, prompting a discussion of emerging trends that developed further into preliminary themes. These discussions included the rest of the research team in weekly and ad-hoc group meetings over the course of two weeks. The first author then reviewed the themes against the coded transcripts to ensure they were representative of the original codes. Following this consistency check, the first author coded the remaining interview transcripts. 

Given the sensitive nature of the study, we anonymized participant responses, and removed all personally identifiable information during transcript review and analysis. All quotes we report have been paraphrased to remove any personally identifiable information while preserving the semantic content and intent of participants’ statements. We retained key linguistic markers (e.g., references to dialect mismatch, idioms) when essential to the analytic claim.

\subsection{Positionality Statement} \label{positionality}
Our research team is based in the U.S. and comprises researchers of many identities. We collectively have experience in HCI, digital health, AI ethics, and clinical psychology. We were cautious that our background in HCI and AI might lead us to a solutionist perspective in this study (i.e., that AI \textit{should} be applied to overcome barriers), which we actively tried to avoid by creating storyboards rather than prototypes and focusing on potential risks of AI in addition to opportunities. The team also includes an immigrant, native Spanish-speaking clinical psychologist who has lived experience, expertise, and clinical experience working with LEP populations, providing insight into the framing and content of the storyboards. Her personal connection with two community organizations aided with recruitment (\autoref{sec:recruitment}), establishing familiarity and trust in our research team for those participants. The interviewing researcher is a second-generation immigrant with professional proficiency in Spanish, helping to establish some trust and familiarity with the participants. However, as this researcher does not come from a similar cultural background as the patient navigators or Spanish-speaking LEP population, this difference may have influenced what participants chose to disclose and what the researcher picked up from participant responses.

\section{Results}
\label{sec:findings}

In this section, we cover our findings that address our two research questions. We start by highlighting how LEP individuals' access to health technologies can be impeded by linguistic (\autoref{subsec:findingsLanguage}), cultural (\autoref{subsec:findingsCulture}), and digital (\autoref{subsec:findingsLiteracy} \& \ref{subsec:findingsPrivacy}) divides. We then describe how patient navigators saw the possible applications (\autoref{subsec:findingsOpp}) and risks (\autoref{subsec:findingsRisks}) of AI tools for LEP individuals. In this section and the following, we use the term \textit{clients} to refer to the Spanish-speaking LEP individuals that the patient navigators work with, reflecting the terminology most commonly used by the navigators themselves. \autoref{tab:findings} summarizes our findings of the barriers that LEP individuals face (RQ1, \autoref{subsec:findingsLanguage}-\autoref{subsec:findingsPrivacy}), and \autoref{tab:findingsRQ2} summarizes our findings of the risks and opportunities for AI to support health communication (RQ2, \autoref{subsec:findingsOpp}-\autoref{subsec:findingsRisks}). Both tables also include associated AI design considerations (\autoref{sec:discussion}).

\begin{table*}[h]
\sffamily
\small
\centering
\caption{Summary of barriers LEP individuals face with respect to health technologies (RQ1), how those barriers manifest, and associated design implications for AI. Bullet points map observed challenges to a potential implication for AI design.}
\begin{tabular}{p{0.16\textwidth} p{0.37\textwidth} p{0.37\textwidth}}
    \textbf{Barrier} & \textbf{Observed Challenges} & \textbf{Implications for AI} \\
    \midrule
    
    Linguistic variations & 
    \begin{minipage}[t]{\linewidth}
        \raggedright
        \begin{itemize}[leftmargin=*, topsep=0pt, itemsep=0pt, parsep=0pt]
            \item Dialectal or regional colloquialisms within a language \vspace{10pt}
            \item Context-sensitive word choices 
            \item Indigenous languages within a Spanish-speaking country \vspace{1mm}
        \end{itemize}\vspace{1mm} 
    \end{minipage} &
    \begin{minipage}[t]{\linewidth}
        \begin{itemize}[leftmargin=*, topsep=0pt, itemsep=0pt, parsep=0pt]
            \item Adapt to dialects and colloquialisms to account for regional and cultural nuances
            \item Align models to individual preferences and mannerisms
            \item Expand models for low-resource languages
        \end{itemize}
    \end{minipage} \\
    \arrayrulecolor{mygray} \hline
     \vspace{-1mm} Cultural differences & 
     \vspace{-1mm} \begin{minipage}[t]{\linewidth}
        \raggedright
        \begin{itemize}[leftmargin=*, topsep=0pt, itemsep=0pt, parsep=0pt]
            \item Traditional or home remedies may conflict with western medical advice
            \item Fear of being judged prevents full disclosure
            \item Respect for authority discourages contradicting or asking questions
        \end{itemize}
    \end{minipage} &
    \vspace{-1mm} \begin{minipage}[t]{\linewidth}
        \raggedright
        \begin{itemize}[leftmargin=*, topsep=0pt, itemsep=0pt, parsep=0pt]
            \item Acknowledge cultural practices and values while offering practical steps for care
            \item Capture nonverbal cues that signify discomfort 
            \item Avoid didactic or dismissive tones that misalign advice by ignoring cultural nuances
        \end{itemize}
    \end{minipage} \vspace{1mm} \\
    \hline
    \vspace{-1mm} Literacy & 
     \vspace{-1mm} \begin{minipage}[t]{\linewidth}
        \raggedright
        \begin{itemize}[leftmargin=*, topsep=0pt, itemsep=0pt, parsep=0pt]
            \item Low reading literacy in English and/or Spanish
            \item Low digital literacy due to limited experience with usable technology (phones, computers, Internet) \vspace{10pt}
            \item Lack of specialized medical knowledge complicates understanding health information
        \end{itemize}
    \end{minipage} &
    \vspace{-1mm} \begin{minipage}[t]{\linewidth}
        \raggedright
        \begin{itemize}[leftmargin=*, topsep=0pt, itemsep=0pt, parsep=0pt]
            \item Offer voice- or picture-based controls
            \item Accompany AI with resources to teach prompting and information validation skills; build simpler models that do not require state-of-the-art devices
            \item Provide simplified explanations that do not require extensive prior medical knowledge
        \end{itemize}
    \end{minipage} \vspace{1mm} \\
    \hline
    \vspace{-1mm} Privacy & 
     \vspace{-1mm} \begin{minipage}[t]{\linewidth}
        \raggedright
        \begin{itemize}[leftmargin=*, topsep=0pt, itemsep=0pt, parsep=0pt]
            \item Reluctance to provide personal information \vspace{10pt}
            \item Fear that personal data could end up in unwanted entities
            \item Hesitancy to engage with digital systems, preferring in-person means
        \end{itemize}
    \end{minipage} &
    \vspace{-1mm} \begin{minipage}[t]{\linewidth}
        \raggedright
        \begin{itemize}[leftmargin=*, topsep=0pt, itemsep=0pt, parsep=0pt]
            \item Access AI services without requiring personal information; anonymize when possible
            \item Minimize storing data and history; offer options to anonymize
            \item Build models that run locally on-device, without requiring external data transmission
        \end{itemize}
    \end{minipage} \vspace{0mm}\\
    \arrayrulecolor{black} \bottomrule
\end{tabular}
\label{tab:findings}
\Description{A table showing 4 rows of barriers that patients face with respect to health technologies. Columns include barriers, observed challenges, and implications for AI.}
\end{table*}

\subsection{Linguistic variation complicated communication surrounding care (RQ1)}
\label{subsec:findingsLanguage}
All patient navigators consistently emphasized that being non-native English speakers imposed communication challenges on clients when accessing care. However, these language barriers included linguistic variation that went beyond the support of translators (human or AI-powered).  

Within native Spanish speakers, dialectal and colloquial differences between various countries or regions sometimes created misunderstandings (P9, P10, P12). Furthermore, merely being from a country where Spanish is the official language did not guarantee that an individual spoke Spanish. Some navigators reported supporting individuals from Guatemala, where the official and most spoken language is Spanish, but where there are also over 20 Mayan languages spoken, such as Q'anjob'al or K'iche'. As a native Spanish speaker from Mexico, P10 recounts an experience while supporting her students: \textit{\begin{quote} "Kids from Guatemala and Honduras have words I’ve never heard before. Sometimes they try to explain it, and it becomes even more confusing. One of the kids described his condition as `falling asleep,' and I was trying to explain seizures, but they have a different word for seizures in his language from Guatemala\ldots Even when we all speak Spanish, we all have different words."
\end{quote}}

This linguistic variation posed challenges for medical interpretation, as most Spanish interpreters that the navigators and their patients encountered were from Mexico and thus less familiar with other dialects or languages. P10 explained that individuals \textit{"speaking Q'anjob'al or other Guatemalan languages often struggle, since phone interpreters use words they can't understand."} Patients recognized the limitations of interpretation services. For example, when asked for language preferences, Guatemalan individuals often default to Spanish despite inevitable misunderstandings because Spanish interpreters were easier to access (P3, P5). 

The context of conversations also played a role in the nuances and connotations of words or expressions. P9 explained that although medical terminology is generally universal, \textit{"words like `stomach', `chest', `shoulder', or `neck' can vary slightly. Sometimes translators use words from different dialects or formalities, which can confuse patients. Small differences can sometimes be the issue for patients."} However, navigators generally agreed that patients still understand the main points even if specific words differ slightly. The dialect mismatches contributed more to friction that reduced their comfort and ease of interaction with the medical system.

\subsection{Cultural differences led to friction with the U.S. healthcare system (RQ1)} \label{subsec:findingsCulture}

Navigators pointed to clients' cultural beliefs and values that clashed with U.S. healthcare practices and western medicine. P10 described her students' distrust of the U.S. health system, saying that \textit{"even students who have been here for many years don't go to the doctor here. They talk to their grandmas or their moms in Guatemala\ldots and their family sends them medicine."} P10 also recalled a particular instance in which the cultural belief of \textit{brujería} (witchcraft) of her student from Guatemala obstructed his opportunity to receive care in the United States: \begin{quote} \textit{"He kept having horrible cramps in his legs at night and spine pain. We went to the doctor, who recommended x-rays and additional tests, but he never wanted to go. He said he knew what he had: a problem caused by a type of demon in Guatemala, and the only cure was to see a} curandero \textit{there. This is more common than I would like to see. Many kids say, `I went to the doctor, and they say I don't have anything, but I know I have it, and they cannot see it because they're not from my country. Only people from my country can see it.'"} \end{quote}

Many clients preferred to follow the medical traditions from their native cultures, which may sometimes be at odds with healthcare practices recommended in the U.S. Clients often refrained from disclosing their home practices to providers of differing backgrounds due to fear of being judged (P5), thus increasing the disconnect between providers and their knowledge of cultural practices. P6 highlighted an example around sleep practices for babies: \begin{quote} \textit{"Doctors recommend cribs, but families often co-sleep. They’ll usually tell the doctor they use cribs because they know it’s the expected answer. If I’m there, I might say, `If they’re co-sleeping, what should they keep in mind?' That way the doctor can offer safer practices."} \end{quote}

The cultural disconnect can be further complicated by a deeply-rooted respect for authority that Hispanic cultures often uphold. P6 explained that \textit{"the doctor is the authority, the translator is the authority, and [the patients] just sit there nodding their head, so they often leave saying they didn't understand."} P10 and P11 echoed nearly identical experiences. The navigators also acknowledged that being proactive necessitated advocating for their clients when communicating with providers and other authority figures. P6 said: \textit{"When I see that it doesn't look like they're understanding something, I prompt them like, `Remember you told me you had a question about this?' So it's become a really big part of what I do."} 

\subsection{Reading and digital literacy limited technology use (RQ1)}
\label{subsec:findingsLiteracy}

Both reading literacy and digital literacy were barriers for many Spanish-speaking individuals in the communities that the patient navigators supported. P11 acknowledged a generation of immigrants that \textit{"had to move to the States in the middle of their schooling, and switching languages mid-education has an impact. Not all of them finished secondary education."} Consequently, most clients who were illiterate could not read English nor Spanish, rendering it nearly impossible to utilize phones or computers without assistance. Navigators agreed, however, that even if their clients were educated or literate, specialized medical knowledge would still pose issues in understanding care information.
 
In some cases, navigators aimed to circumvent reading literacy issues via technology accessibility features such as sending audio messages rather than text or touching pictures to navigate (P6). However, these practices sometimes collided with clients' lack of digital literacy. P6 described an unexpected challenge when accommodating Guatemalan families with low reading literacy: \begin{quote} \textit{"I created a flyer with pictures and a QR code that they could scan and hear the information in Q'anjob'al, but we learned that a lot of families didn't know how to use a QR code. I thought this was a really easy way for them to be able to hear information in their native language, but then there was another barrier."} \end{quote}

Digital literacy varied greatly across LEP individuals. Navigators attributed generational differences to the range of familiarity with technology: \textit{"Most adults never had to use a computer before. Families here for many years already have teenage or adult children who handle technology, so adults don't feel the need\ldots even to use email. However, new immigrants, here for 5-10 years, force themselves to learn technology because even job applications are online now"} (P10). P8 stated that most of his adult clients use flip phones or landlines and do not feel the need to use a smartphone, even if given one. On the other hand, navigators that worked with teenagers and young adults explained that their younger clients are comfortable with technology. P10, a school district family liaison, explained that it is easy for teenagers that had never used a phone or computer before immigrating to the U.S. to adopt these technologies effectively. 

Low digital literacy often stemmed from barriers to accessing technology in the first place, creating additional sources of stress and confusion when seeking medical care. Clients of P13, a social worker in a major metropolitan area, often \textit{"use old phones\ldots which are now considered obsolete. They run into connection problems, apps that don’t work well, or just really slow devices. Patients sometimes give up. That adds stress - on top of being sick, now they're also struggling with technology."} P3 echoed this sentiment, citing financial burdens as a barrier to accessing usable phones or apps. Several navigators said very few of their clients owned laptops or computers, and P6 and P14 said most of their clients did not have stable internet at home (P6's clients lived in trailer parks and had to go outside the trailer to get better phone service) and instead relied on mobile hotspots. P4 noticed that most individuals do not even know about new technology, highlighting a lack of educational resources in the community. This unfamiliarity with technology creates discomfort: P12 described that one of the students in her digital literacy class \textit{"was so nervous, scared to even touch the computer. She didn't want to break it."} Barriers to access and low digital literacy caused patients to struggle with even small tasks that seemed trivial even to navigators (P6). For example, patient navigators needed to show their clients how to join Zoom meetings (P6), maneuver the cursor on computers (P12), and overcome frustration with applications that require verification or extra steps (P13). 

At the same time, LEP individuals recognized the importance of technology for accessing care. P3 realized that his clients \textit{"understand the importance of investing in technology."} P12 and P14, program coordinators of a digital literacy program within a social services organization that serves Spanish-speaking individuals, aimed to increase educational resources and access to technology. They taught individuals in the community how use computers, AI, and the internet to search for information. P14 said that if given opportunities and support to overcome access barriers, \textit{"students and community members genuinely want to learn—they have the desire to improve."} All navigators agreed that with proper support and guidance, AI has the potential to improve the wellbeing of their clients by increasing on-demand information access and translations for language support.

\subsection{Privacy concerns further discourage use of technology (RQ1)}
\label{subsec:findingsPrivacy}
Independent of literacy issues, LEP individuals had heightened distrust of digital privacy policies and therefore hesitation to use technology, being afraid that their data would be used wrongly or end up in the possession of unintended entities (P8, P10). While nearly all navigators agreed that patients were generally wary of sharing personal data, one offered a contrasting view: P7, a school district parent liaison, reasoned that most of the families he works with cannot even read well in Spanish, so ethical concerns rarely cross their minds. However, some navigators drew examples from other applications that underscored their clients' concern for guarding their personal data. P5 worked with clients who \textit{"keep money in a shoebox. If they don’t trust banks or the government, they surely won’t ask [AI] for help."} P6 worked with families who refuse to even obtain driver's licenses: \begin{quote}\textit{"They say, `No, I’ll be in the system, [government officials] will find me.' It’s a very real fear that impacts their lives and decisions in big ways. They know people who’ve been stopped without a license, they know the fees, the risks of police and court. And that’s still preferable to them compared to being in the system."}\end{quote} These concerns stem primarily from administrative policies and immigration status (P3, P5, P12, P13), even for first-generation immigrants who have relatives or friends that are undocumented (P9). Many navigators themselves have experienced recently growing distrust from their own clients, despite already having established familiarity and trust. This reluctance to share personal information extends to patients' unwillingness to seek healthcare services, which usually require patients to disclose personal data (P8). Lack of familiarity with technology also created distrust: \textit{"[Clients] feel uneasy when websites ask for personal information. A lot of it comes from lack of knowledge—they don’t understand what a chatbot is, so they assume it’s a robot trying to steal their information"} (P14).

Interestingly, navigators drew distinctions between types of data that their clients would be comfortable sharing with clinics and technology. P14 outlined the difference as: \textit{"They’re very protective of sensitive information like Social Security numbers or bank accounts. But they’re comfortable sharing details about feelings, medical issues, recipes, or business ideas—things that can’t be stolen or hacked."} Other navigators generally agreed, drawing the line between \textit{"willing to give symptoms and medical history and maybe an email"} but \textit{"hesitant with phone numbers, addresses, or other personally identifying information"} (P6). These boundaries also extended to AI, as most navigators reasoned that clients would be comfortable sharing their symptoms with AI to get information, but would be less willing to log in to an account and save their chat history.

Navigators employed various strategies to assuage their clients' privacy concerns. P4 emphasized to her clients the confidential nature of their interactions, informing them of their rights to take action or bring her to justice if their information is shared with any outside sources. P3, P8, P10, P12, and P13 employed similar strategies but also noted the trade-offs: as P3 explained, \textit{"We need to sometimes reassure them that\ldots the state won’t share their information. The forms even say the information is confidential. Sometimes we tell them they’re not required to give certain information, so many clients end up providing only very basic information."} P10 emphasizes that her clients need reassurance that their information would not be shared with anyone else, leveraging the personal relationship and trust she established with her clients in order to obtain the information needed to support them. 

\begin{table*}[h]
\sffamily
\small
\centering
\caption{Summary of risks and opportunities for AI health technologies to facilitate communication (RQ2) and associated design implications for AI. Bullet points map observed behavior to a potential implication for AI design.}
\begin{tabular}{p{0.01\textwidth} | >{\raggedright\arraybackslash}p{0.14\textwidth} p{0.37\textwidth} p{0.37\textwidth}}
     & \textbf{Theme} & \textbf{Observed Behavior} & \textbf{Implications for AI} \\
    \midrule
    
    \multirow{2}{*}{\parbox[c][13\baselineskip][c]{\linewidth}{\centering\rotatebox{90}{Opportunities}}} & Reduce social barriers &
    \begin{minipage}[t]{\linewidth}
        \raggedright
        \begin{itemize}[leftmargin=*, topsep=0pt, itemsep=0pt, parsep=0pt]
            \item Patients feel unheard by providers due to language or cultural misunderstandings
            \item Patients feel embarrassed or uncomfortable when describing medical conditions
            \item Telehealth distance emphasizes alienation
        \end{itemize}
    \end{minipage} &
    \begin{minipage}[t]{\linewidth}
        \begin{itemize}[leftmargin=*, topsep=0pt, itemsep=0pt, parsep=0pt]
            \item Provide on-demand conversational support that eliminates social pressures of a human-human conversation
            \item Detect confusion or jargon in real-time conversations and provide simplified explanations
            \item Provide on-demand support by allowing patients to externalize their sentiments \vspace{2mm}
        \end{itemize}
    \end{minipage} \\
    \arrayrulecolor{mygray} \cline{2-4}
     &  \vspace{-1mm} Alleviate resource constraints &
      \vspace{-1mm} \begin{minipage}[t]{\linewidth}
        \raggedright
        \begin{itemize}[leftmargin=*, topsep=0pt, itemsep=0pt, parsep=0pt]
            \item Long wait times for interpreters
            \item Navigators are overburdened to properly follow-up with patients 
            \item Long work hours increase risk of navigator errors and decreased service quality
        \end{itemize}
    \end{minipage} &
     \vspace{-1mm}\begin{minipage}[t]{\linewidth}
        \raggedright
        \begin{itemize}[leftmargin=*, topsep=0pt, itemsep=0pt, parsep=0pt]
            \item Provide on-demand translation
            \item Proactively send check-in questionnaires to monitor patient progress
            \item Automate navigator work, identify errors, and ensure consistency
        \end{itemize} \vspace{1mm}
    \end{minipage} \\
    \arrayrulecolor{black}\midrule
    \multirow{2}{*}{\parbox[c][8\baselineskip][c]{\linewidth}{\centering\rotatebox{90}{Risks}}} & Loss of Human Connection &
     \begin{minipage}[t]{\linewidth}
        \raggedright
        \begin{itemize}[leftmargin=*, topsep=0pt, itemsep=0pt, parsep=0pt]
            \item Human presence is significant source of comfort
            \item Patients want to feel that someone cares
        \end{itemize}
    \end{minipage} &
    \begin{minipage}[t]{\linewidth}
        \raggedright
        \begin{itemize}[leftmargin=*, topsep=0pt, itemsep=0pt, parsep=0pt]
            \item Keep navigators in the loop to use AI alongside patients
            \item Use AI to support rather than replace patient-navigator relationships\vspace{1mm}
        \end{itemize}
    \end{minipage} \\
    \arrayrulecolor{mygray} \cline{2-4}
     & \vspace{-1mm}Misinformation &
     \vspace{-1mm}\begin{minipage}[t]{\linewidth}
        \raggedright
        \begin{itemize}[leftmargin=*, topsep=0pt, itemsep=0pt, parsep=0pt]
            \item Patients may not be able to validate AI information or translations 
            \item Patients may overrely on AI-generated information
        \end{itemize}
    \end{minipage} &
    \vspace{-1mm} \begin{minipage}[t]{\linewidth}
        \raggedright
        \begin{itemize}[leftmargin=*, topsep=0pt, itemsep=0pt, parsep=0pt]
            \item Build in accuracy checks, safety filters, and plain-language summaries
            \item Provide educational supports to teach skills around assessing trustworthiness \vspace{1mm}
        \end{itemize}
    \end{minipage} \\
    \arrayrulecolor{black} \bottomrule
\end{tabular}
\label{tab:findingsRQ2}
\Description{A table showing 4 rows of opportunities and risks of AI health technologies. Columns include theme, observed behavior, and implications for AI.}
\end{table*}

\subsection{Opportunities for AI: reducing social barriers and institutional constraints (RQ2)}
\label{subsec:findingsOpp}
Having characterized barriers shaping LEP patients’ technology experiences (RQ1), we now describe how navigators envision AI playing a role in communication dynamics and care workflows. Navigators identified two ways that AI systems would be uniquely positioned to support LEP individuals seeking care: reducing sentiments of alienation and alleviating resource constraints. 

Navigators described that many of their clients felt unheard, rejected, or embarrassed by the healthcare system. P1 explains that \textit{"in Hispanic culture, people like to talk, share details, and give background. Appointments are usually rushed, so doctors usually want `yes/no' answers about symptoms, but patients may tell their whole life story first. That mismatch in communication style can definitely be a barrier."} Other patients left appointments feeling like their concerns went unheard due to language or cultural misunderstandings that impeded their ability and confidence to advocate for themselves (P5, P11). Oftentimes institutional restrictions on interpreters prevented them from addressing these issues. At P6's clinic, the interpreters must strictly provide word-for-word translations, barred from intervening when they perceive misunderstandings between patients and providers. Patients also experienced embarrassment or discomfort when describing medical conditions, especially if doing so in English, causing the patients to refrain from elaborating on their conditions (P2). Lastly, telehealth methods, albeit convenient, may exacerbate these communication disconnects. P5 lamented that telehealth \textit{"make[s] it hard because you can't see the patient to gauge their pain, and Wi-Fi issues make it worse."} Collectively, these barriers caused patients to feel alienated and not seek care: \textit{"Because English isn’t their first language, they’re extra cautious. They feel vulnerable and worry about being scammed. They know they’re a group people might take advantage of, so they’re wary"} (P14). Patient navigators proposed that some of these issues could be alleviated by AI. For example, P2 reasoned that talking to a chatbot about a medical condition would eliminate the sense of embarrassment or social judgment that resides in a human-to-human conversation. P6 proposed the idea of an AI interface that transcribes patient-provider conversations in real-time and detects potentially confusing concepts or jargon in the conversation, allowing patients to gain clarification in real-time without causing embarrassment by verbally interrupting or admitting their confusion. P11 reasoned that talking to AI as a post-visit debrief could help patients feel better by externalizing their sentiments and receiving comfort. 

Navigators also identified ways that AI systems could alleviate institutional burdens and resource constraints. Most clinics and emergency rooms are frequently stretched thin, forcing patients to endure long wait times for an interpreter or a provider, or even reschedule their appointments completely. Navigators agreed that AI could help fill the interpretation gaps by serving as an on-demand translator for when human interpreters are not readily available, thus alleviating wait times and frustration for LEP patients. Navigators whose work required communicating large volumes of complex information to clients noted that AI could help identify errors and ensure consistency, especially when fatigue set in from long work hours (P3). Some navigators also cited not having enough bandwidth to routinely follow-up with clients, such as to ensure correct understanding and adherence to care plans (P13). P9 proposed an AI system that proactively sends check-in questionnaires to patients on a regular basis to help monitor their progress and health without relying on scheduling official visits. This regular communication would also serve as an intervention mechanism to prevent patients from waiting until they are very sick to seek help (P10).

\subsection{Risks of AI: loss of human connection and validation strategies (RQ2)}
\label{subsec:findingsRisks}

While navigators pointed to opportunities for AI to enhance LEP patient care, they also raised concerns about the uniquely-human support that many patients need. Nearly all of the navigators emphasized the significance of human presence as a source of comfort and encouragement for individuals while navigating healthcare, difficult circumstances, or loneliness. P10 described her personal journey battling a long-term medical condition: \begin{quote}
\textit{"When I was at appointments, I was hearing new, scary information. What was really helpful was having someone next to me---a nurse who was very caring---who kept asking if I wanted to stop, if I understood, if I wanted water, if she could help me in any way. That human aspect is very important, especially when hearing scary news. In complex cases, having a human who shows care and presence is important."}
\end{quote}

P14 gave the examples that his clients would rather hand a physical resume to a person in an office, rather than submitting it online, and they would rather have someone personally explain confusing documents such as a privacy policy, rather than asking for explanations from AI. He highlighted that \textit{"it's the personable aspect that matters. They need to feel like someone cares,"} which is conveyed through the personal physical interactions. Navigators worried that these needs would at best make clients uncomfortable with AI and at worse remove a key asset of human support. P5, a community navigator from Mexico, explained: \begin{quote}\textit{"An AI might know the facts, but it’s not Mexican, didn’t grow up in Mexico, so it doesn’t really understand me the way a person would. For example, Mexican remedy traditions—American doctors don’t get them, but other people [from similar cultures] do. AI knows a lot about an egg cleanse, but it hasn’t experienced it, so it doesn’t really know why we do it."}\end{quote} 

Navigators also anticipated risks with misinformation and mistranslations from AI. Most navigators experienced inaccurate information or translations themselves when they used AI in their work, but they worried that their clients would not have the same capabilities to validate outputs. P9 contended that AI translation systems \textit{"really only work for bilingual people. I've had students submit translations without knowing how to adjust words depending on the context."} Other navigators reported similar experiences with AI performing poorly when translating idioms or words in specific contexts. On issues of misinformation, P10 asserted that her clients \textit{"don't ensure information online is accurate. Many trust social media for important information. For example, if TikTok said the president would give green cards to everyone, they believe it. They don’t know how to verify information yet."} Similarly, P11 emphasized medical misinformation from social media and word-of-mouth as a significant issue: \begin{quote}\textit{"Patients often lecture me about diets or remedies they saw on Facebook. Sometimes it’s legitimate, sometimes it’s not. Maybe it’s not immediately harmful, but it complicates care. I’ve had patients stop blood pressure medication after hearing from friends it causes dizziness. That was a known side effect we had already explained. So then we have to re-educate, rebuild trust, and sometimes switch them to another medication just to remove the stigma around the original drug name. It complicates care - not necessarily bad, but challenging."}\end{quote}

Navigators, then, extrapolate these concerns to AI as an information source. Some called for digital literacy education that includes building skills in validating information accuracy, assessing trustworthiness of information sources, navigating online culture, and learning cybersecurity.

\section{Discussion} \label{sec:discussion}

In this study, patient navigators saw clear potential for AI technologies to improve healthcare access and quality for LEP individuals. However, we avoid framing AI as a solution to structural inequities, as successful introduction of AI depends on how well these systems integrate into these existing interpersonal (e.g., patient-navigator or patient-provider) and sociotechnical ecosystems. Thus, in this section, we discuss opportunities and risks for developing AI tools that facilitate communication between LEP individuals, providers, and navigators. For many patients, trust in and consequent adoption of a system is inseparable from the policies and institutions that govern it~\cite{bahari2024trust}, meaning that even well-designed technologies can fail if they do not address broader systemic inequities. Thus, we discuss the implications of our study within these broader structural contexts. We do not generalize the following discussion and considerations to all LEP or even all Spanish-speaking LEP individuals. 

\subsection{Design considerations for embedding AI technology into existing sociotechnical infrastructures} \label{subsec:disc1}

Prior research in care-centered computing advocates for integrating health technologies within existing platforms or systems familiar to patient users ~\cite{anaya2021meeting, regenbrecht2014field, curtis2025use} to minimize burdens and disruptions of adapting to new systems. Consistent with this principle, navigators in our study discussed the difficulties for LEP individuals to adopt new technologies, skills, or even visit a new clinic, as many LEP individuals rely heavily on navigators, family members, or other sources of support to mediate communication, explain processes, or advocate on their behalf. Therefore, rather than demanding users migrate to completely new platforms, we propose integrating AI according to the following principles.

\paragraph{Platform congruence.} To minimize burdens of learning new practices, AI should be deployed on devices and platforms that LEP individuals already use. Navigators explained that their clients were generally more comfortable and had more access to phones (as opposed to computers, laptops, wearables, or other forms of technology). Therefore, AI tools should be mobile-friendly or even mobile-centric. Going further, chatbots could send reminders, answer questions, or translate information along familiar channels, such as SMS or WhatsApp~\cite{unger2018short, 10.1145/3628096.3628752}. This \textbf{"meet them where they are" approach} removes the cognitive load of navigating new interfaces or devices, allowing new AI to integrate quietly in established technological practices.

\paragraph{Small Language Models.} AI must recognize limitations imposed by the digital divide and institutional distrust. Our findings highlight a privacy paradox in which AI requires personal data or third party APIs to tailor generated responses, yet LEP populations are often highly reticent to share their information due to distrust in institutional policies and fearing surveillance. Here, the technical constraints of the digital divide offer a solution to the psychological constraints of trust. By employing either small language models (SLMs) that have fewer parameters or models that can be publicly downloaded (e.g., the Olmo3 model family \citep{olmo2025olmo3} or gpt-oss \citep{agarwal2025gpt}) and run locally, developers can simultaneously address two problems. Models that can run on a mobile device \citep{10.1145/3662006.3662059} can function without stable internet, addressing P6's observation of families relying on intermittent hotspots in mobile trailers. Psychologically, on-device processing ensures data sovereignty such that because data stays on device, the risk of interception or institutional misuse is minimized~\cite{dubiel2024device}. While such models might sacrifice some compute-intensive capabilities (e.g. deep reasoning~\cite{openai2025deep}), allowing patients to access tools locally may be more effective than policy statements in reassuring privacy. Furthermore, giving users agency over their data through options to use nicknames, ephemeral chat modes, or privacy-enhancing systems~\cite{pham2025proxygptenablinguseranonymity} can further align the system with user needs for privacy. However, on-device processing does not eliminate all risk (e.g., device loss, coercive access, or organizational policies around tool use), so privacy protections must also include usability-centered safeguards (e.g., quick-delete, no default history, clear threat models). AI design should also explicitly consider risks such as inadvertent disclosure to family members, clinic staff, or authorities; device sharing; and fear of surveillance tied to immigration enforcement. These threats shift design priorities toward minimal data retention, non-identifying access modes, and clear user control over what is stored or shared~\cite{li2024human}.

\paragraph{Enforcing trust and appropriate power dynamics.} Prior work established that cultural values may influence acceptance and choice of technology~\cite{harrington2022s, 10.1145/1970378.1970382} due to factors such as alignment of communication style, sensitivity in presenting information, and readability of information~\cite{10.1145/3706598.3713362}. For instance, navigators emphasized that high respect for authority in Hispanic cultures rendered patients unwilling to fully disclose their symptoms. Therefore, in order to encourage comfortable and transparent interactions, AI systems should adopt the tone of an intermediary between medical authority and accessible peer by integrating computational approaches to politeness~\cite{danescu-niculescu-mizil-etal-2013-computational} or style transfer~\cite{reif-etal-2022-recipe}. In this way, AI can serve as a \textbf{supportive scaffold that reinforces rather than usurps the established relationships} of trust between patients and navigators. A practical pattern is “coach and translator”: the system helps patients rehearse questions in their preferred dialect, flags jargon, and generates respectful question prompts (e.g., “Can you explain that in another way?”) while explicitly encouraging patients to defer medical decisions to clinicians.

\subsection{AI-driven personalization to support long tail linguistic, cultural, and literacy variation}
\label{subsec:disc3}

\paragraph{Linguistic variation.}
Languages and cultures are not monolithic. In our study, navigators highlighted how dialectal differences can cause misunderstandings, even if interpretation is available. Furthermore, individuals from predominantly Spanish-speaking countries did not necessarily speak Spanish. Though recent years have seen improvements in the multi-lingual capabilities of LLMs~\cite{downey2024targeted, huang2025benchmax}, our findings suggest AI systems should \textbf{prioritize dialect-level adaptation}, ensuring that they can flexibly account for regional, cultural, and individual nuances~\cite{capdevila2025crossing}. However, existing frontier models remain limited in recognizing rich diatopic variability, including in languages like Spanish~\cite{mayor2025s,tellez2023regionalized}, highlighting the need for benchmark datasets that can rigorously evaluate LLM capabilities in understanding and responding to diverse dialects~\cite{martinez2025spanish,grandury2025leaderboard}. Personalization here requires systems to align with individual communication preferences, which can be achieved via few-shot learning~\cite{kim-yang-2025-shot} or other alignment methods~\cite{shaikh2024show, wu2024aligning} that offer avenues for AI to continually learn the patient's vocabulary and preferences.

\paragraph{Cultural variation.} While health practices are deeply rooted in cultural care routines, navigators explained that patients often conceal cultural practices such as egg cleanses from providers due to fear of judgment or failure to understand. While providers currently use questionnaires and wearable technology~\cite{bharmal2023health, mak2025integrating} to capture patient lifestyle and background, raw data alone rarely translates into cultural empathy. AI tools have potential to synthesize \textbf{culturally-appropriate insights grounded in lived experiences} for providers~\cite{peters2025towards}. One way of achieving this is to pair a generalist high-level LLM and a group of culturally-tuned LLMs in order to incorporate diverse cultural perspectives~\cite{sorensen2024roadmap} and thus validate medical advice against specialized cultural norms. Furthermore, in-context adaption via sociodemographic prompting or cultural role-play~\cite{shaikh2023modeling, hwang2023aligning} can allow the agent to proactively identify and bridge value gaps between patients and providers. In this way, an AI system could, for example, detect hesitance regarding a treatment plan and generate culturally relevant analogies or explanations that align with the patient’s lived experience~\cite{peters2025towards}. However, this requires a shift from static profiling to dynamic cultural alignment, ensuring that systems avoid stereotyping by continuously learning from the user’s specific feedback and value signals~\cite{10.1145/3706598.3713477, wu-etal-2025-aligning}.

Conversely, for patient-facing interactions, AI systems should \textbf{recognize how cultural beliefs around medicine intersect with biomedical recommendations}, framing guidance in a way that is both culturally sensitive and clinically appropriate. Value-sensitive design suggests that systems should acknowledge the user’s lived experiences before pivoting to clinical corrections to maintain trust and adherence~\cite{Liao2024AI}. In situations where following cultural beliefs could lead to serious health consequences, such as the case of P10's student who believed \textit{brujería} (witchcraft) caused his excruciating leg and spine pain, AI should be governed by context-aware safety alignment~\cite{shen2025valuecompass} to gently prioritize biomedical advice, ensuring that critical interventions are not overlooked. AI can minimize the risk of alienating patients during critical moments by framing medical interventions as necessary complements to, rather than replacements for, their cultural practices.

\paragraph{Literacy variation.} Assumptions of baseline literacy or technological familiarity often exclude marginalized users from health technologies or general digital services~\cite{yao, alkureishi, whitehead}. Literacy, however, is multi-dimensional, encompassing reading, digital, health, and even specialized forms of AI literacy such as scientific and data literacy~\cite{long2020ai}. Navigators emphasized that LEP individuals often struggle concurrently with multiple dimensions of literacy such that an AI tool that simplifies text to a 5th-grade level is still inaccessible since many patients lack the skills to type queries or navigate interfaces, often relying on other modalities such as voice messages to communicate. Therefore, AI systems should \textbf{adapt content and modality simultaneously to address multi-dimensional literacy gaps}. The complexity of content can be dynamically adjusted by providing definitions of unfamiliar terms, plain-language summaries~\cite{10.1145/3589955}, or simplifications based on a user's demonstrated reading~\cite{naeem2025eduadapt} or health literacy level~\cite{schillinger2021employing}. Furthermore, AI interfaces for LEP populations should be multimodal by default. Given the dual threads of agents that can navigate complex digital environments~\cite{shaikh2025creating, Mozannar2025MagenticUITH, Huq2025CowPilotAF} and generative voice-based interfaces that can overcome reading and typing barriers~\cite{pradhan, le2023voicebox}, there is an opportunity to develop voice-based agents that can, for example, help patients navigate other tools (e.g., Zoom, digital calendars, and email). Voice-based translation agents~\cite{hudelson2024using} may also support the frequent over-the-phone interactions between interpreters, patients, and providers. Furthermore, text-to-image models can generate visual explanations~\cite{wu-etal-2025-healthcards} such as diagrams showing how to take medication, tailored to the specific medication or dosage. Leveraging generative AI capabilities to move beyond text-centric interventions can help bridge the multifaceted literacy gaps faced by LEP individuals.

\subsection{When AI might not be the best approach}\label{subsec:disc4}
Our study highlighted that while AI systems hold potential to support LEP individuals seeking care, they may not always be the most appropriate or effective solution. In some contexts, introducing complex technological ecosystems may create more friction than strengthening existing or deploying low-tech strategies.

Navigators in our study echoed the call in research communities for technology that complements, rather than replaces, human expertise in healthcare settings~\cite{alrassi2021technology, sezgin2023artificial}. However, our results highlight a tension: while most navigators were optimistic that the benefits of AI for LEP individuals would outweigh risks, \textbf{successful deployment of AI places the burden on navigators to teach and monitor their clients} who may be unfamiliar with or unwilling to adopt new technologies due to barriers in literacy and access to hardware. Thus, LEP adoption of AI creates a form of invisible labor~\cite{10.1023/A:1008694621289}, which navigators expressed their willingness to shoulder because increasing digital literacy and overall education empowers patients to advocate for themselves. However, because human-in-the-loop systems would offload burdens of error checking and training on these already-stretched navigators, designing AI for this ecosystem may require intermediaries~\cite{10.1145/3449118} specifically for navigators.

Many communication needs could also be adequately addressed through offline resources rather than technology. Static resources (e.g., FAQs, worksheets, pamphlets) that are helpful for conveying important health information are traditionally written by providers or navigators. Rather than attempting to replace these materials, \textbf{AI could instead play a supporting role} by, for example, generating these resources that are then refined by humans. These resources could then be distributed with minimal digital access as printed sheets or shared audio files~\cite{verran2024artificial}. This hybrid workflow allows providers and navigators to leverage the generative power of AI without imposing the privacy risks and usability hurdles of LEP patient-AI interaction~\cite{toyama2015geek}.

Introducing AI necessitates AI literacy that goes beyond basic digital skills~\cite{long2020ai}. Navigators described already investing significant effort in teaching their clients foundational digital skills such as using computers, sending emails, and logging into online patient portals. Learning additional skills such as prompt engineering or information validation increases the barrier to entry for patients to adopt AI safely and productively~\cite{10.1145/3706598.3713841}. When the \textbf{cognitive load of learning the tool may outweigh immediate utility}, investments in improving existing workflows, interpreter access, or patient education may create more immediate impact as a first step rather than deploying new AI tools.

\subsection{Limitations and Future Work}
\label{sec:limitations}

While the storyboard scenarios provided a useful structure for discussion, they reflect hypothetical scenarios, thus capturing patient navigators' expectations and attitudes rather than observed behaviors. Furthermore, the storyboards may have guided participants toward particular ideas or interpretations. Nevertheless, because all but one of the interviewed navigators found the scenarios to be realistic, they were able to relate the fictional scenarios to their actual lived experiences, thus grounding the discussions in real examples that allowed them to critically reflect on how AI might affect their existing practices and relationships. Future work should incorporate in-situ or observational methods such as participatory design sessions or diary studies that capture actual behaviors and decision-making processes.

Our findings are also context-specific and do not necessarily generalize to other LEP populations or even Spanish-speaking populations in different regions of the U.S. Cultural, linguistic, and systemic characteristics of any particular area may shape the needs and experiences of those patients and navigators in ways that may differ from our study context. Therefore, while navigators can generalize across patients in their own regions and cultural backgrounds, they cannot speak for LEP individuals that they do not regularly interact with. Additionally, because this study captures only the perspectives of navigators, which we selected as an early-phase, risk-sensitive step into LEP communities and cultural norms before talking to patients directly, this unilaterial perspective raises two limitations. First, their position as intermediaries between patients and providers may bias their perceptions. For instance, while they may have concerns of AI around potential job displacement, all navigators emphasized aspects of their roles that were irreplaceable, such as driving patients to the hospital or teaching them how to use online portals. Nonetheless, their responses may still be inherently influenced by their positionality as navigators. Second, since these insights reflect secondhand interpretations rather than firsthand experiences of patients themselves, triangulating findings with provider and LEP patient perspectives would capture a more holistic narrative. Navigators may overestimate patients’ willingness to adopt AI tools or understate concerns that patients do not disclose to intermediaries. Therefore, future work should validate these themes through participatory sessions with LEP patients and providers. Building prototypes or low-fidelity design probes informed by the findings of this study, such as voice-based interfaces, could ground the participatory discussions with patients while offering insight into the validity and extensions of our findings. In this way, we can ensure that AI systems are built to be responsive and effective to the needs of the communities they intend to serve.
\section{Conclusion}

In this study, we interviewed 14 patient navigators who support Spanish-speaking LEP individuals in receiving healthcare. We identified key sociotechnical and cultural factors that shape the way LEP individuals access health technologies and communicate with providers. Our findings reveal linguistic, cultural, literacy, and privacy barriers that LEP individuals face when interacting with healthcare services and technologies. These findings motivate opportunities for AI to be situated within existing practices and limitations, while considering issues of data privacy, misinformation, and cultural alignment. Future work should solicit the perspectives of LEP individuals to directly inform AI system designs that best support marginalized communities in receiving healthcare.

\bibliographystyle{ACM-Reference-Format}
\balance
\bibliography{00-main}

\newpage
\appendix
\section{AI Literacy Questionnaire} \label{ailiteracy}

In the recruitment form, we adapted a basic AI literacy questionnaire from~\cite{grassini2024psychometric}. All questions were asked on a 5-point Likert scale ranging from 1 (strongly disagree) to 5 (strongly agree). The questions were as follows, along with the mean and standard deviation of the participant responses:

\begin{enumerate}
    \item I understand the basic concepts of artificial intelligence. $(M = 3.86, SD = 0.86)$
    \item I can judge the pros and cons of AI. $(M = 3.64, SD = 0.74)$
    \item I keep up with the latest AI trends. $(M=2.86, SD=1.03)$
    \item I am comfortable talking about AI with others. $(M=3.86, SD=1.23)$
    \item I can think of new ways to use existing AI tools. $(M=3.36, SD=1.08)$
\end{enumerate}

\subsection{Storyboards}\label{app:storyboards}

\begin{figure*}[h!]
 \begin{subfigure}[b]{\columnwidth}
    \centering
\includegraphics[width=0.9\columnwidth]{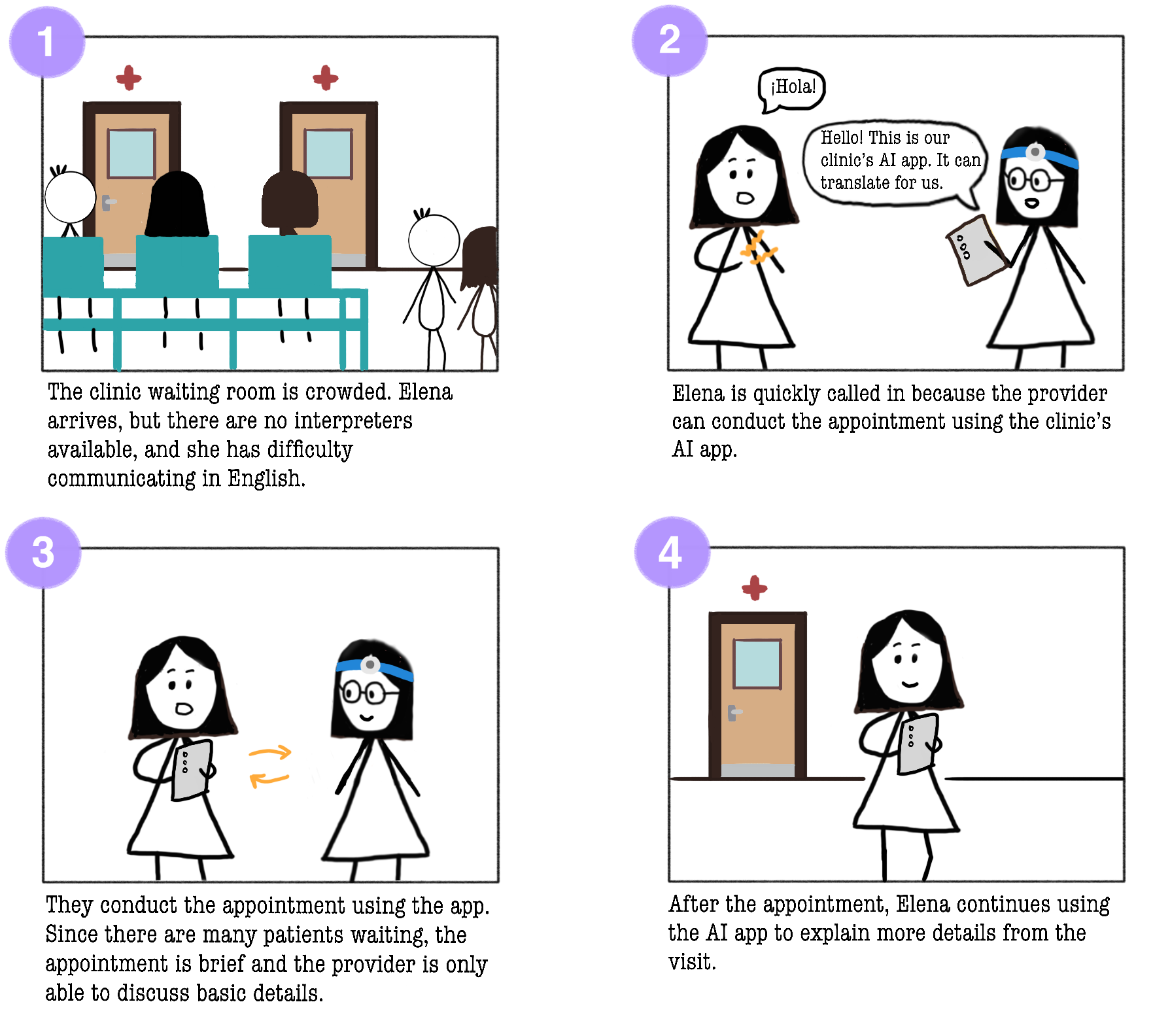}
    \caption{Clinical Integration \& Support}
    \label{fig:s1}
    \end{subfigure}\hfill
\begin{subfigure}[b]{\columnwidth}
    \centering
\includegraphics[width=0.9\columnwidth]{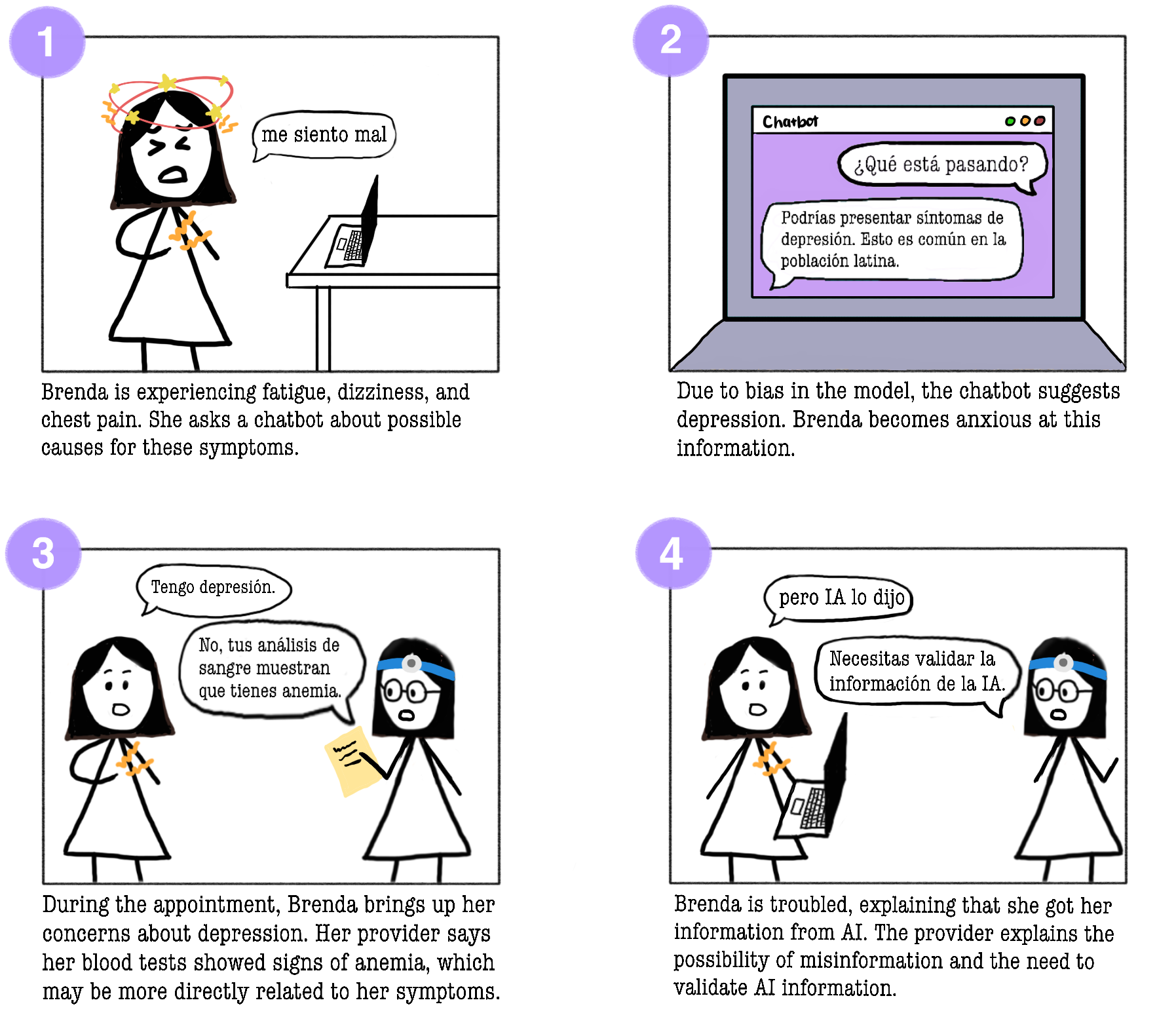}
    \caption{Clinical Safety \& Accuracy}
    \label{fig:s2}
    \end{subfigure}\hfill 
\begin{subfigure}[b]{\columnwidth}
    \centering  \includegraphics[width=0.9\columnwidth]{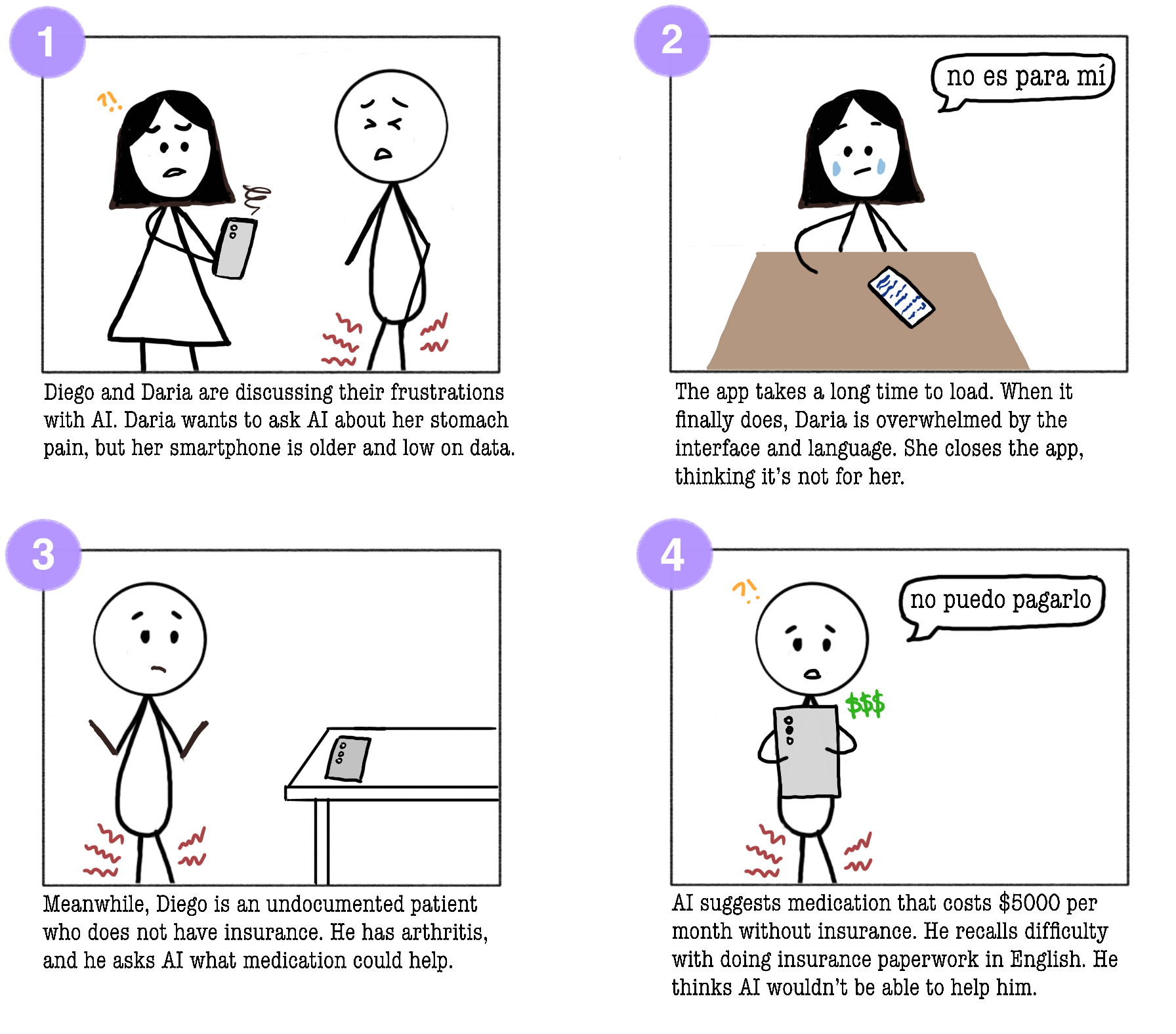}
    \caption{Fairness, Equity, \& Access}
    \label{fig:s3}
    \end{subfigure}\hfill
\begin{subfigure}[b]{\columnwidth}
    \centering
\includegraphics[width=0.9\columnwidth]{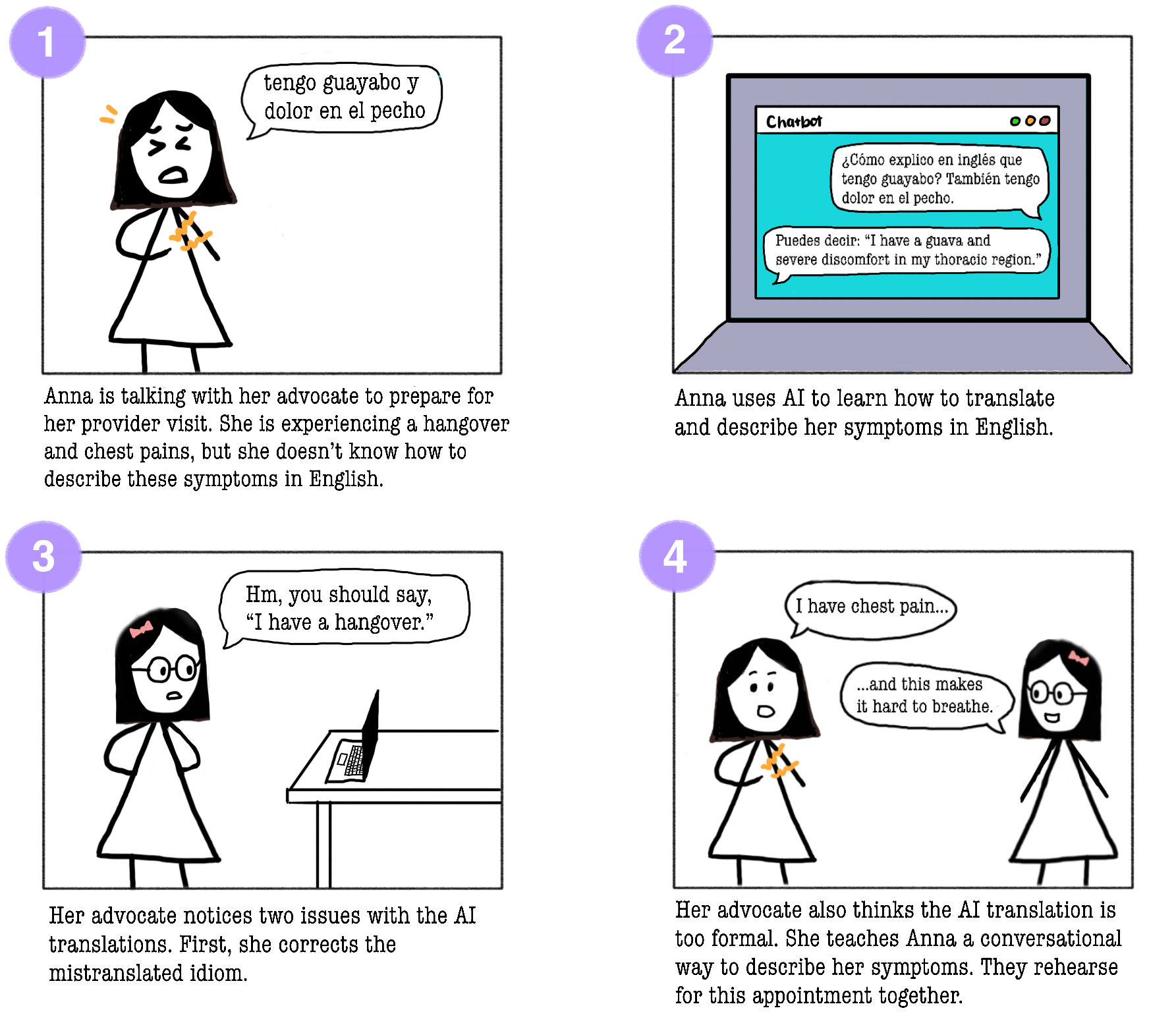}
    \caption{Language \& Communication}
    \label{fig:s4}
    \end{subfigure}\hfill 
\begin{subfigure}[b]{\columnwidth}
    \centering
\includegraphics[width=0.9\columnwidth]{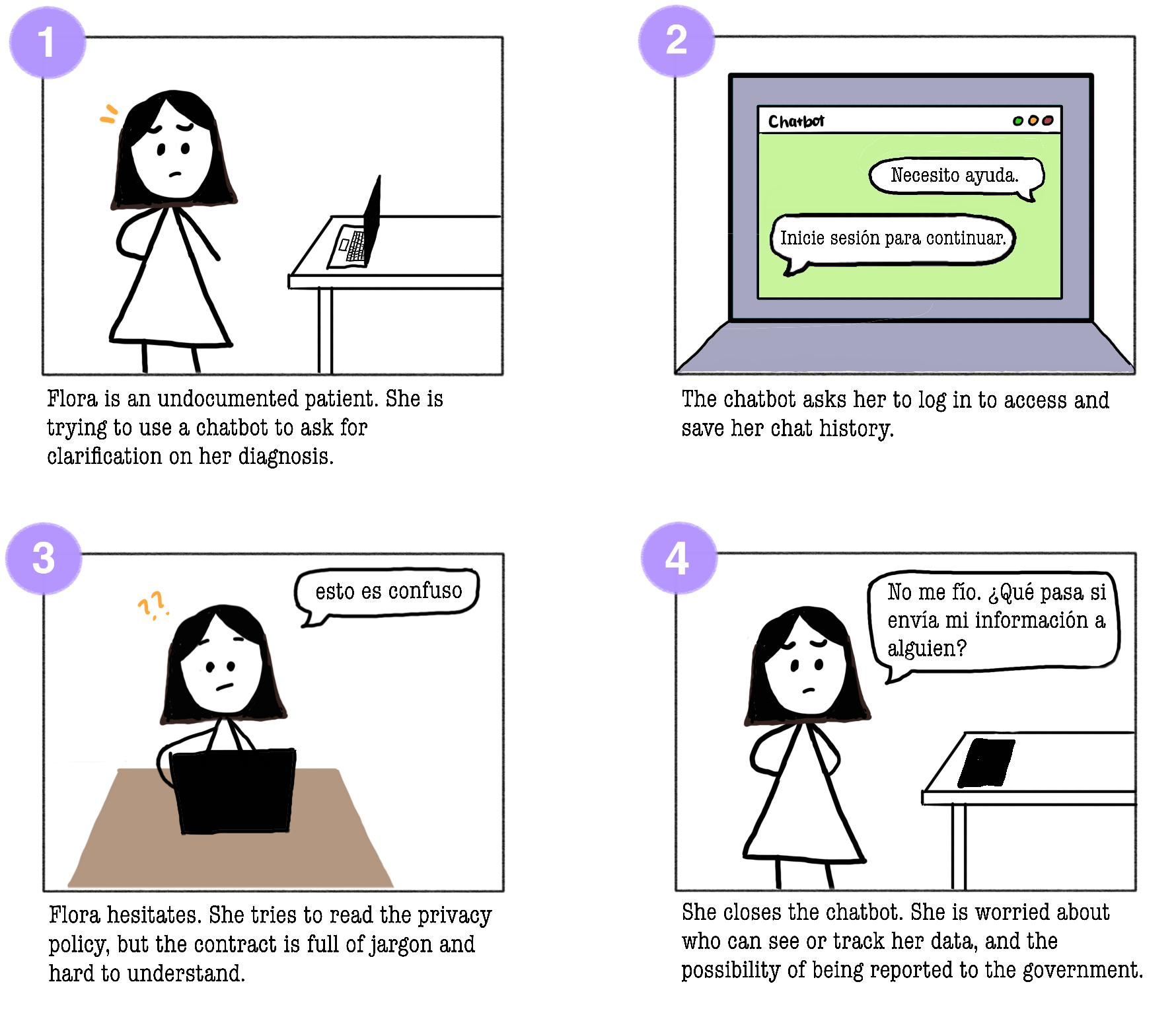}
    \caption{Privacy, Transparency, \& Ethics}
    \label{fig:s5}
    \end{subfigure}\hfill
\begin{subfigure}[b]{\columnwidth}
    \centering
\includegraphics[width=0.9\columnwidth]{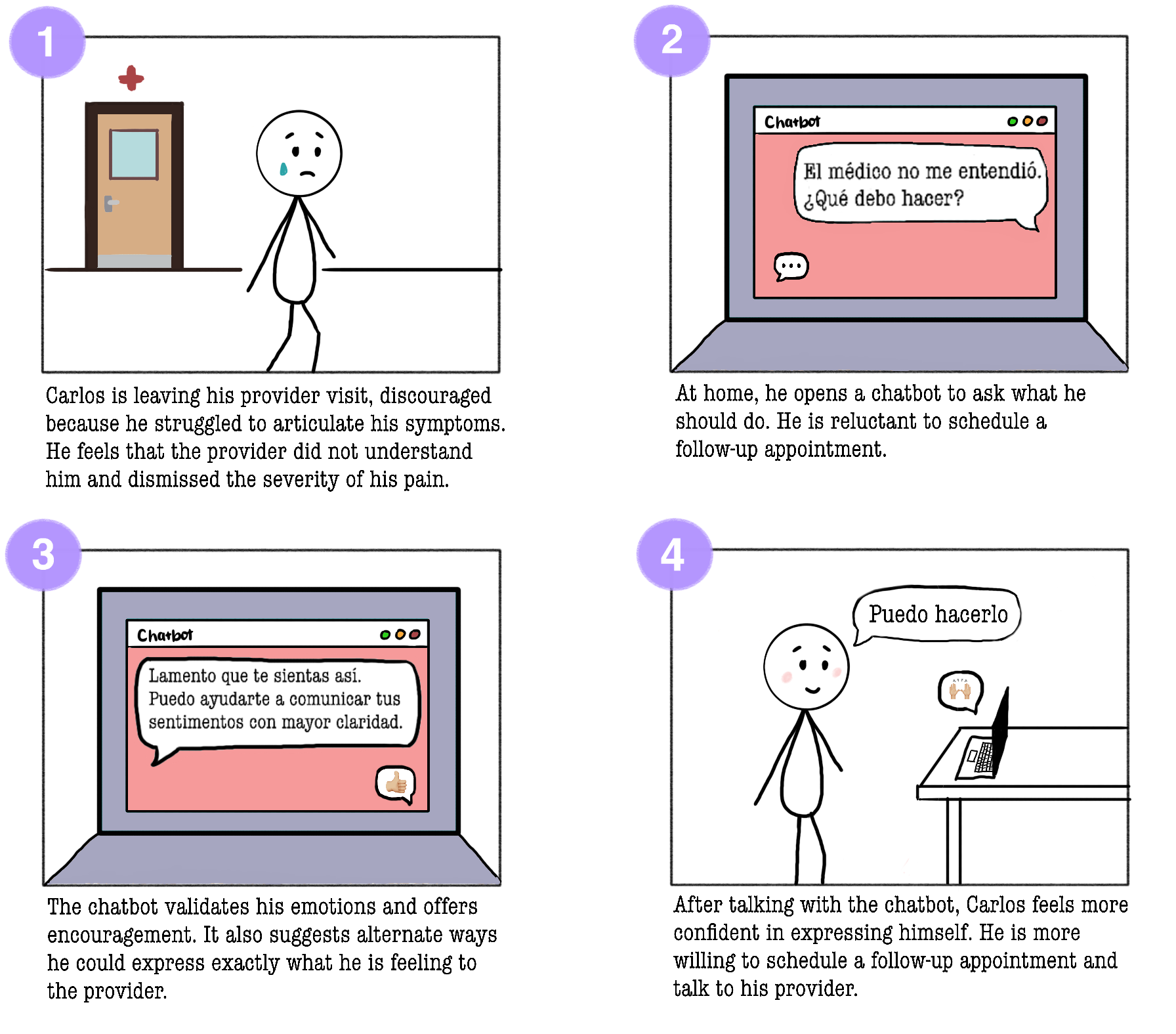}
    \caption{Relationality \& Comfort}
    \label{fig:s6}
    \end{subfigure}\hfill   
    \caption{Storyboards used in interviews with patient navigators.}
    \label{fig:all_storyboards}
    \Description[storyboards]{6 illustrated storyboards each depicting a theme as described in Methods.}
\end{figure*}

We include the list of all the storyboards in our study in~\autoref{fig:all_storyboards}.

\end{document}